\documentclass[10pt,journal,compsoc]{IEEEtran}
\ifCLASSOPTIONcompsoc
  \usepackage[nocompress]{cite}
\else
  \usepackage{cite}
\fi
\ifCLASSINFOpdf
  \usepackage[pdftex]{graphicx}
\else
\fi
\usepackage{amsmath}
\usepackage{array}

\usepackage[hidelinks]{hyperref}
\usepackage{pdfpages}
\usepackage{textcomp}

\ifCLASSOPTIONcompsoc
\usepackage[caption=false,font=footnotesize,labelfont=sf,textfont=sf]{subfig}
\else
\usepackage[caption=false,font=footnotesize]{subfig}
\fi
\begin{document}
%
\title{Convolutional Neural Networks for\\Spectroscopic Redshift Estimation\\on Euclid Data}
%
%
%
%

\author{\IEEEauthorblockN{Radamanthys Stivaktakis$^{1,2}$, Grigorios Tsagkatakis$^2$, Bruno Moraes$^3$, \\ Filipe Abdalla$^{3,4}$, Jean-Luc Starck$^5$, Panagiotis Tsakalides$^{1,2}$\\}
\IEEEauthorblockA{Department of Computer Science - University of Crete, Greece$^1$\\}
\and
\IEEEauthorblockA{Institute of Computer Science - Foundation for Research and Technology (FORTH), Greece$^2$\\}
\and
\IEEEauthorblockA{Department of Physics \& Astronomy, University College London, UK$^3$\\}
\and
\IEEEauthorblockA{Department of Physics and Electronics, Rhodes University, South Africa$^4$\\}
\and
\IEEEauthorblockA{Astrophysics Department - CEA Saclay, Paris, France$^5$\\}}

\IEEEpubid{This work has been submitted to the IEEE for possible publication. Copyright may be transferred without notice, after which this version may no longer be accessible.}


\IEEEtitleabstractindextext{%
\begin{abstract}
In this paper, we address the problem of spectroscopic redshift estimation in Astronomy. Due to the expansion of the Universe, galaxies recede from each other on average. This movement causes the emitted electromagnetic waves to shift from the blue part of the spectrum to the red part, due to the Doppler effect. Redshift is one of the most important observables in Astronomy, allowing the measurement of galaxy distances. Several sources of noise render the estimation process far from trivial, especially in the low signal-to-noise regime of many astrophysical observations. In recent years, new approaches for a reliable and automated estimation methodology have been sought out, in order to minimize our reliance on currently popular techniques that heavily involve human intervention. The fulfilment of this task has evolved into a grave necessity, in conjunction with the insatiable generation of immense amounts of astronomical data. In our work, we introduce a novel approach based on Deep Convolutional Neural Networks. The proposed methodology is extensively evaluated on a spectroscopic dataset of full spectral energy galaxy distributions, modelled after the upcoming Euclid satellite galaxy survey. Experimental analysis on observations of idealistic and realistic conditions demonstrate the potent capabilities of the proposed scheme.
\end{abstract}

\begin{IEEEkeywords}
Astrophysics, Cosmology, Deep Learning, Convolutional Neural Networks, Spectroscopic Redshift Estimation, Euclid.
\end{IEEEkeywords}}

\maketitle

\IEEEdisplaynontitleabstractindextext

%
\IEEEpeerreviewmaketitle

\IEEEraisesectionheading{\section{Introduction}\label{sec:introduction}}

%
%
%
%
%
\IEEEPARstart{M}{odern} cosmological and astrophysical research seeks answers to questions like ``what is the distribution of dark matter and dark energy in the Universe?'' \cite{bertone2010particle, DarkEnergy2006}, or ``how can we quantify transient phenomena, like exoplanets orbiting distant stars?'' \cite{marcy2005observed}. To answer such questions, a large number of deep space observation platforms have been deployed. Spaceborne instruments, such as the Planck Satellite\footnote{http://www.esa.int/Our\textunderscore Activities/Space\textunderscore Science/Planck} \cite{Planck2015Cosmo}, the Kepler Space Observatory\footnote{http://kepler.nasa.gov/} \cite{borucki2010kepler} and the upcoming Euclid mission\footnote{http://sci.esa.int/euclid/}\cite{laureijs2011euclid}, seek to address these questions with unprecedented  accuracy, since they avoid the deleterious effects of Earth's atmosphere, a strong limiting factor to all their observational strategies. Meanwhile, ground-based telescopes like the LSST\footnote{https://www.lsst.org} \cite{abell2009lsst} will be able to acquire massive amounts of data through high frequency full-sky surveys, providing complementary observations. The number and capabilities of cutting-edge scientific instruments in these and other cases have led to the emergence of the concept of Big Data \cite{bryant_bigdata}, mandating the need for new approaches on massive data processing and management. The analysis of huge numbers of observations from various sources has opened new horizons in scientific research, and astronomy is an indicative scenario where observations propel the data-driven scientific research \cite{stephens2015big}.

\begin{figure*}[!t]
\subfloat[Clean Rest-Frame Spectral Profile]{\includegraphics[height=4cm,width=5.3cm]{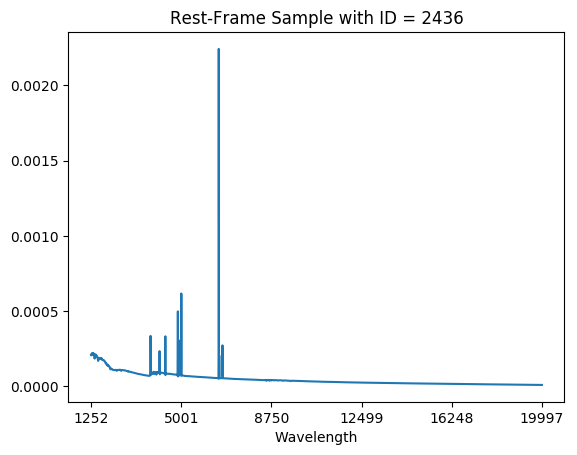}}
\hspace*{1cm}
\subfloat[Clean (Randomly) Redshifted Equivalent]{\includegraphics[height=4cm,width=5.3cm]{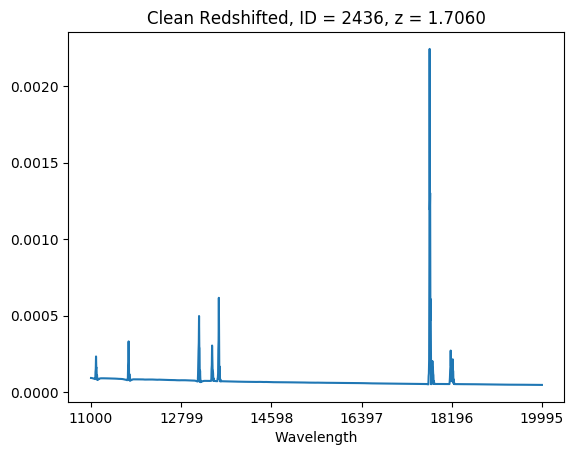}}
\hspace*{1cm}
\subfloat[Noisy Redshifted Equivalent]{\includegraphics[height=4cm,width=5.3cm]{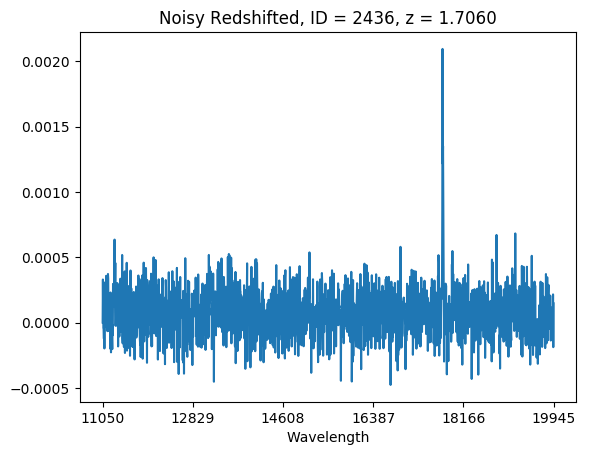}}
\vspace*{7mm}
\caption{Examples of the data used. From the initial rest-frame samples, we produce random redshifted samples in clean and noisy forms. The y-axis corresponds to the spectral density flux value, in a normalized form.\newline}
\label{examples}
\end{figure*}

One particular long-standing problem in astrophysics is the ability to derive precise estimates to galaxy redshifts. According to the Big Bang model, due to the expansion of the Universe and its statistical homogeneity and isotropy, galaxies move away from each other and any given observation point. A result of this motion is that light emitted from galaxies is shifted towards larger wavelengths through the Doppler effect, a process termed \textit{redshifting}. 
Redshift estimation has been an integral part of observational cosmology, since it is the principal way in which we can measure galaxies' radial distances and hence their 3-dimensional position in the Universe. This information is fundamental for several observational probes in cosmology, such as the rate of expansion of the Universe and the gravitational lensing of light by the matter distribution - which is used to infer the total dark matter density - among other methods\cite{efstathiou1990cosmological,massey2010dark}.

The Euclid satellite aims to measure the global properties of the Universe to an unprecedented accuracy, with emphasis on a better understanding of the nature of Dark Energy. It will collect \textit{photometric} data with broadband optical and near-infrared filters and \textit{spectroscopic} data with a near-infrared slitless spectrograph. The latter will be one of the biggest upcoming spectroscopic surveys, and will help us determine the details of cosmic acceleration through measurements of the distribution of matter in cosmic structures. In particular, it will measure the characteristic distance scale imprinted by primordial plasma oscillations in the galaxy distribution. The projected launch date is set for 2020 and throughout its 6-year mission, Euclid will gather of the order of 50 million galaxy spectral profiles, originating from wide and deep sub-surveys. A top-priority issue associated with Euclid is the efficient processing and management of these enormous amounts of data, with scientific specialists from both astrophysical and engineering backgrounds contributing to the ongoing research. To successfully achieve this purpose, we need to ensure that realistically simulated data will be available, strictly modeled after the real observations coming from Euclid in terms of quality, veracity and volume.

Estimation of redshift from spectroscopic observations is far from straightforward. There are several sources of astrophysical and instrumental errors, such as readout noise from CCDs, contaminating light from dust enveloping our own galaxy, Poisson noise from photon counts, and more. Furthermore, due to the need of obtaining large amounts of spectra, astronomers are forced to limit the time of integration for any given galaxy, resulting in low signal-to-noise measurements. As a consequence, not only it becomes difficult to confidently measure specific spectral features for secure redshift estimation, but we also incur the risk of misidentifying features - e.g. confusing a hydrogen line for an oxygen line - which results in so-called catastrophic outliers. Human evaluation mitigates a lot of these problems with current - relatively small - data sets. However, Euclid observations will be particularly challenging, working in very low signal-to-noise regimes and obtaining a massive amount of spectra, which will force us to develop automated methods capable of high accuracy and necessitating minimal human intervention.

Meanwhile, the rise of the ``golden age" of Deep Learning \cite{lecun2015deep} has fundamentally changed the way we handle and apprehend raw, unprocessed data. While the existing machine learning models heavily rely on the development of efficient feature extractors, a task non-trivial and very challenging, Deep Learning architectures are able to single-handedly derive important characteristics from the data by learning intermediate representations and by structuring different levels of abstraction, essentially modelling the way the human brain works. The monumental success of Deep Learning networks in recent years, has been strongly enhanced by their interminable capacity to harness the power of Big Data and fully exploit emerging, cutting-edge hardware technologies, constituting one of the currently most widely used paradigms in numerous applications and in various scientific research fields.

One such a network subsists in Convolutional Neural Networks (CNNs) \cite{lecun_lenet}, a sequential model structured with a combination of Convolutional \& Non-Linear Layers. The inspiration behind Convolutional Neural Networks resides in the concept of visual receptive fields \cite{hartline1938response}, i.e. the region in the visual sensory periphery where stimuli can modify the response of a neuron. This is the main reason that CNNs initially found application in image classification, by learning to recognize images by experience, in the same perception where a human being can gradually learn to distinguish different image stimuli from one another. Today, CNNs are administered in the use of various types of data, with more or less complicated dimensional structures, with the key property of maintaining their spatial correlations without the need to collapse higher dimensional matrices into flattened vectors.

Our main motivation lies in the use of a state-of-the-art model, such as Convolutional Neural Networks, for an automated and reliable solution of the problem of spectroscopic redshift estimation. Estimating galaxy redshifts is perceived as a regression procedure, still a classification approach can be formulated without the loss of essential information. The robustness of the proposed model will be examined in two different data variations, as depicted in the example of Figure \ref{examples}. In the first case (b), we deploy randomly redshifted variations of the original rest-frame spectral profiles of the dataset used, substantially constituting linear translations of the rest-frame, in logarithmic scale. This is considered an idealistic scenario, as it ignores the interference of noise or presumes the existence of a reliable denoising technique. On the other hand, a more realistic case (c) is studied, with the available redshifted observations subjected to noise of realistic conditions.

The main contributions of our work are referenced below:

\begin{itemize}

\item We use a Deep Learning architecture for the case of spectroscopic redshift estimation, never used before for the issue at hand. To achieve that we need to convert the problem from a regression task, as engaged in general, to a classification task, as encountered in this novel approach.\newline

\item We utilize Big Data and evaluate the impact of a significant increase of the employed observations in the overall performance of the proposed methodology. The dataset used is modelled after one of the biggest upcoming spectroscopic surveys, the Euclid Mission \cite{laureijs2011euclid}.\newline

\end{itemize}

The outline of this paper is structured as follows. In Section \ref{SoTA}, we overview the related work in redshift estimation and Convolutional Neural Networks in general. In Section \ref{Proposed}, we describe 1-Dimensional CNNs and we analyse the formulated methodology. In Section \ref{ADeeperPerspective}, we mainly focus on the dataset used and describe its properties. In Section \ref{Experimental}, we present the experimental results, with accompanying discussion. Conclusion and future work are engaged in Section \ref{Conclusions}.


\section{Related Work}\label{SoTA}

Photometric observations have been extensively utilized in redshift estimation due to the fact that photometric analysis is substantially less costly and time-consuming, contrary to the spectroscopic case. 
Popular methods for this kind of estimation include Bayesian estimation with predefined spectral templates \cite{benitez_bayes}, or alternatively some widely used machine-learning models, adapted for this kind of problem, like the Multilayer Perceptron \cite{bonnett_nn}, \cite{sadeh_annz} and Boosted Decision Trees \cite{sadeh_annz}, \cite{gerdes_arborz}. However, the limited wavelength resolution of photometry, compared to spectroscopy, introduces a higher level of uncertainty to the given procedures. In spectroscopy, by observing the full Spectral Energy Distribution (SED) of a galaxy, one can easily detect distinctive emission and absorption lines that can lead to a judicious redshift estimation, by measuring the wavelength shift of these spectral characteristics from the rest frame. Due to noisy observations, the main redshift estimation methods involve cross-correlating the SED with predefined spectral templates\cite{glazebrook_cross} or PCA decompisitions of a template library. Noisy conditions and potential errors due to the choice of templates are the main reason that most reliable spectroscopic redshift estimation methods heavily depend on human judgment and experience to validate automated results.

The existing Deep Learning models (i.e. Deep Artificial Neural Networks - DANNs) have largely benefited from the dawn of the Big Data era, being able to produce impressive results, that can match, or even exceed, human performance \cite{goodfellow_dlbook}. Despite the fact that training a DANN can be fairly computationally demanding as we increase its complexity and the data it needs to process, nevertheless, the rapid advancements on computational means and memory storage capacity have rendered feasible such a task. Also, contrary to the training process, the final estimation phase for a large set of testing examples can be exceptionally fast, with an execution time that can be considered trivial. Currently, Deep Learning is considered to be the state-of-the-art in many research fields, such as image classification, natural language processing and robotic control, with models like  Convolutional Neural Networks \cite{lecun_lenet}, Long-Short Term Memory (LSTM) networks \cite{hochreiter1997long}, and Recurrent Neural Networks \cite{hopfield1987neural}, dominating the research field.

The main idea behind Convolutional Neural Networks materialized for the first time with the concept of ``Neocognitron'', a hierarchical neural network capable of visual pattern recognition \cite{fukushima_neo}, and evolved into LeNet-5, by Yann LeCun et al. \cite{lecun_lenet}, in the following years. The massive breakthrough of CNNs (and Deep Learning in general) transpired in 2012, in the ImageNet competition \cite{imagenet_2009}, where the CNN of Alex Krizhevsky et al. \cite{krizhevsky_2012}, managed to reduce the classification error record by \texttildelow{}10\%, an astounding improvement at the time. CNNs have been considered in numerous applications, including image classification \cite{krizhevsky_2012} \cite{simonyan2014very} \& processing \cite{zagoruyko2015learning}, video analytics \cite{tsagkatakis2017goal} \cite{karpathy2014large}, spectral imaging \cite{fotiadou2017deep} and remote sensing \cite{hu2015transferring} \cite{hu2015deep}, confirming their dominance and ubiquity in contemporary scientific research. In recent years, the practice of CNNs in astrophysical data analysis has led to new breakthroughs, among others, in the study of galaxy morphological measurements and structural profiling through their surface's brightness \cite{tuccillo2016deep} \cite{tuccillo2017deep}, the classification of radio galaxies \cite{aniyan2017classifying}, astrophysical transients \cite{gieseke2017convolutional} and star-galaxy seperation \cite{kim2016star}, and the statistical analysis of matter distribution for the detection of massive galaxy clusters, known as strong gravitational lenses \cite{petrillo2017finding} \cite{lanusse2017cmu}. The exponential increase of incoming data, for future and ongoing surveys, has led to a compelling need for the deployment of automated methods for large-scale galaxy decomposition and feature extraction, negating the commitment on human visual inspection and hand-made user-defined parameter setup.

\section{Proposed Methodology}\label{Proposed}

\begin{figure*}[!t]
\centering
\includegraphics[width=5in]{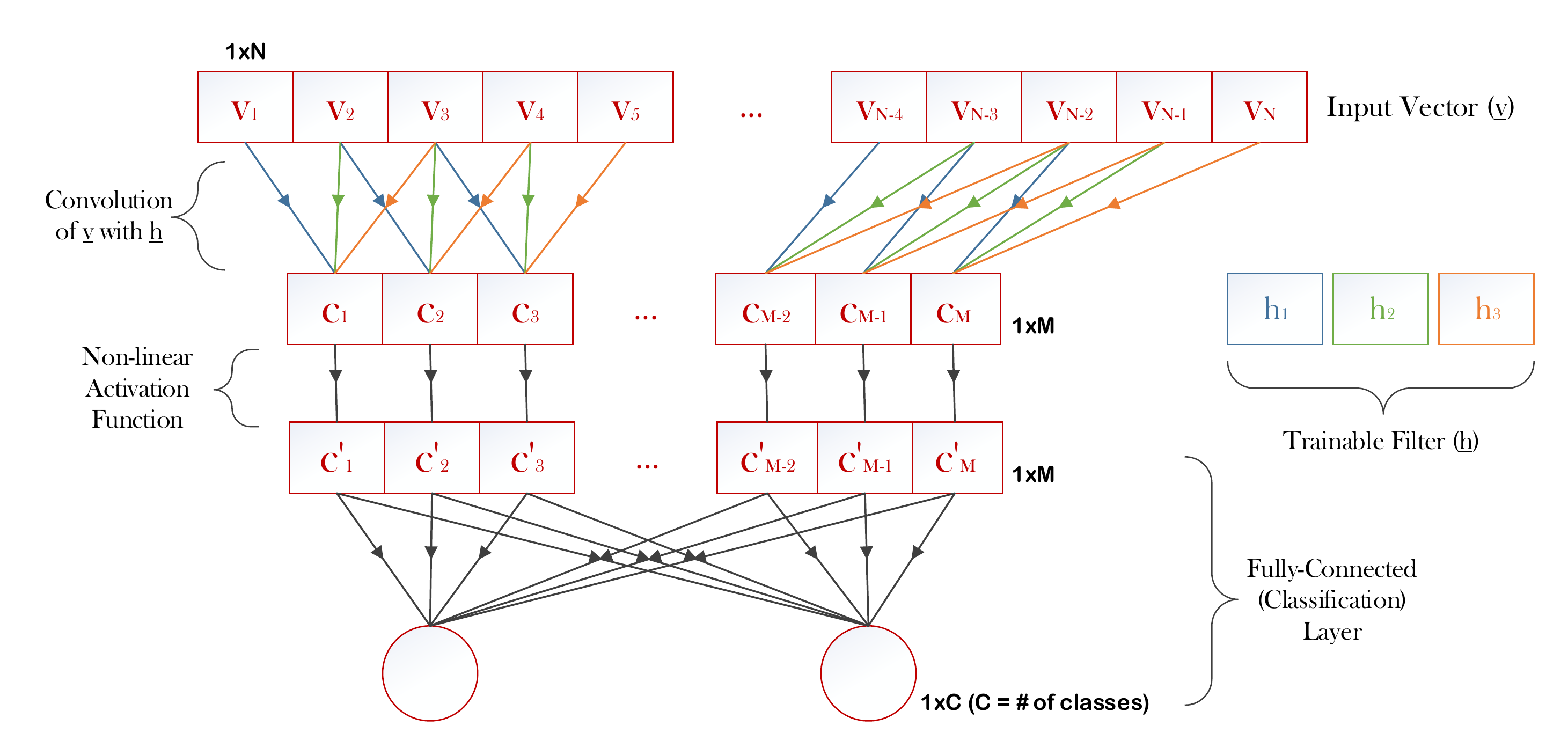}
\caption{Simple 1-Dimensional CNN. The input vector \underline{v} is convolved with a trainable filter \underline{h} (with a stride equal to 1), resulting in an output vector of size $M=N-2$. Subsequently, a non-linear transfer function (typically ReLU) is applied element-wise on the output vector preserving its original size. Finally, a fully-connected, supervised layer is used for the task of classification. The number of the output neurons (C) is equal to to the number of the distinct classes of the formulated problem (800 classes in our case).}
\label{simple_cnn}
\end{figure*}

In this work, we study the problem of accurate redshift estimation from realistic spectroscopic observations, modeled after Euclid. Redshift estimation is considered to be a regression task, given the fact that a galaxy redshift (\textit{z}) can be measured as a non-negative real valued number (with zero corresponding to the rest-frame). Given the specific characteristics of Euclid, we can focus our study in the redshift range of detectable galaxies. Subsequently, we can restrict the precision of each of our estimations to match the resolution of the spectroscopic instrument, meaning that we can split the chosen redshift range into evenly sized slots equal to Euclid's required resolution. Hence, we can transform the problem at hand from a regression task to a classification task using a set of ordinal classes, with each class corresponding to a different slot, and accordingly we can utilize a classification model (Convolutional Neural Networks in our case) instead of a regression algorithm.

\subsection{Convolutional Neural Networks}

A Convolutional Neural Network is a particular type of Artificial Neural Network, which comprises of inputs, outputs and intermediate neurons, along with their respective connections, which encode the learnable weights of the network. One of the key differences between CNNs and other neuronal architectures, like Multilayer Perceptron \cite{rosenblatt1961principles}, is that in typical neural networks, each neuron of a given layer connects with every neuron of its respective previous and following layers (fully-connected layers) contrary to the CNN case, where the aforementioned network is structured in a locally-connected manner. This local-connectivity property exhibits spatial correlations of the given data with the assumption that neighboring regions of each data-example are more likely to be related than regions that are farther away. By reducing the number of total connections, we manage to dramatically decrease, at the same time, the number of trainable parameters, rendering the network less prone to overfitting. 

\subsubsection{Typical Architecture of a 1-Dimensional CNN}

A typical 1D CNN (Figure \ref{simple_cnn}) is structured in a sequential manner, layer by layer, using a variety of different layer types. The foundational layer of a CNN is the \textit{Convolutional Layer}. Given an input vector of size (1xN) and a trainable filter (1xK), the convolution of the two entities will result in a new output vector with a size (1xM), where $M=N-K+1$. The value of M may vary based on the stride of the operation of convolution, with bigger strides leading to smaller outputs. In the entirety of this paper, we assume the generic case of a stride value equal to 1.

The trainable parameters of the network (incorporated in the filter), are initialized randomly \cite{bengio_grad} and, therefore, are totally unreliable, but as the training of the network advances, through the process of backpropagation \cite{rumelhart_backprop}, they are essentially optimized and are able to capture interesting features from the given inputs. The parameters (i.e. weights) of a certain filter are considered to be shared \cite{lecun_shared}, in the aspect that the same weights can be used throughout the convolution of the entirety of the input. This, can consequently lead to a drastical decrease in the number of weights, enhancing the ability of the network to generalize and adding to its total robustness against overfitting. To ensure that all different features can be captured in the process, more than one filters can be actually used.

In more difficult problems, using one Convolutional Layer is insufficient, if we want to construct a reliable and more complex solution. A deeper architecture, able to derive more detailed characteristics from the training examples, is a necessity. To cope with this issue, a non-linear function can be interjected between adjacent Convolutional Layers, enabling the network to act as a universal function approximator \cite{hornik_approx}. Typical choices for the non-linear function (known as \textit{activation function}) include the logistic (sigmoid) function, the hyperbolic tangent (tanh) and the Rectified Linear Unit (ReLU). The most common choice in CNNs is ReLU ($f(x) = max(0,x)$) and its variations \cite{xu_relu}. Compared to the cases of the sigmoid and hyperbolic tangent functions, the rectifier possesses the advantage that it is easier to compute (as well as its gradient) and is resistant to saturation conditions \cite{krizhevsky_2012}, rendering the training process much faster and less likely to suffer from the problem of vanishing gradients \cite{hochreiter1998vanishing}.

Finally, one or more \textit{Fully-Connected Layers} are typically introduced as the final layers of the CNN, committed to the task of the supervised classification. A Fully-Connected Layer is the typical layer met in Multilayer Perceptron and as the name implies, all its neuronal nodes are connected with all the neurons of the previous layer leading to a very dense connectivity. Given the fact that the output neurons of the CNN correspond to the unique classes of the selected problem, each of these neurons must have a complete view of the highest-order features extracted by the deepest Convolutional Layer, meaning that they must be necessarily associated with each of these features.

The final classification step is performed using the multi-class generalization of Logistic Regression known as \textit{Softmax Regression}. Softmax Regression is based on the exploitation of the probabilistic characteristics of the normalized exponential (softmax) function below:

\begin{equation}
h_{\theta}(x)_{j} = \frac{e^{\theta_{j}^{T}x}}{\sum_{k=1}^{W} e^{\theta_{k}^{T}x}},
\end{equation}

where x is the input of the Fully-Connected Layer, $\theta_{j}$ are the parameters that correspond to a certain class $w_{j}$ and W is the total number of the distinct classes related to the problem at hand. It is fairly obvious that the softmax function reflects an estimation of the normalized probability of each class $w_{j}$, to be predicted as the correct class. As deduced from the previous equation, each of these probabilities can take values in the range of [0,1] and obviously, they all add up to the value of 1. This probabilistic approach composes a good reason for the transformation of the examined problem to a classification task, rendering possible to quantify the level of confidence for each estimation and providing a clearer view on what has been misconstrued in the case of misclassification.

The use of \textit{Pooling Layers} has been excluded from the pipeline, given the fact that pooling is considered, among others, a great method of rendering the network invariant to small changes of the initial input. This is a very important property in image classification, but in our case these translations of the original rest-frame SEDs, almost define the different redshifted states. By using pooling, we suppress these transformations, ``crippling" the network's ability to identify each different redshift.\newline

\subsubsection{Regularizing Techniques}

In very complex models, like ANNs, there is always the risk of overfitting the training data, meaning that the network produces over-optimistic predictions throughout the training process, but fails to generalize well on new data, subsequently leading to a decaying performance. The local neuronal connectivity that is employed in Convolutional Neural Networks, and the concept of weight sharing, reported in the previous paragraphs, cannot suffice in our case, given the fact that the single, final Fully-Connected Layer (which contains the majority of the parameters) will consist of hundreds of neurons.
One way to address the problem of the network's high variance exists in the use of Big Data, with a theoretical total negation of the effects of overfitting, when the number of training observations tend to infinity. We will thoroughly examine the impact of the use on Big Data, on clean and noisy observations, in our experimental scenarios.

\textit{Dropout} \cite{srivastava2014dropout} and \textit{Batch Normalization} \cite{ioffe2015batch} are, also, two very popular techniques in CNNs that can help narrow down the consequences of overfitting. In Dropout, the following simple, yet very powerful trick is used to temporarily decrease the total parameters of the network at each training iteration. All the neurons in the network are associated with a probability value \textit{p} (subject to hyper-parameter tuning) and each neuron, independently from the others, can be dropped from the network (along with all incoming and outgoing connections) with that probability. Bigger values for \textit{p} lead to a more degenerated network, while, on the other hand, lower values affect in a more ``lightweight" way its structure. Each layer can be associated with a different \textit{p} value, meaning that Dropout can be considered as a per-layer operation with some layers discarding neurons in a higher percentage, while others dropping neurons in a lower rate or not at all. In the testing phase, the entirety of the network is used, meaning that Dropout is not applied at all.

Batch Normalization, on the other hand, can be accounted for, more as a normalizer, but previous studies \cite{ioffe2015batch} have shown that it can work very effectively as a regularizer as well. Batch Normalization is, in fact, a local (per layer) normalizer, that operates on the neuronal activations in a way similar to the initial normalizing technique applied to the input data in the pre-processing step. The primary goal is to enforce a zero mean and a standard deviation of one, for all activations of the given layer and for each mini-batch. The main intuition behind Batch Normalization lies in the fact that, as the neural network deepens, it becomes more probable that the neuronal activations of intermediate layers might diverge significantly from desirable values and might tend towards saturation. This is known as Internal Covariate Shift \cite{ioffe2015batch} and Batch Normalization can play a crucial role on mitigating its effects. Consequently, it can actuate the gradient descent operation to a faster convergence \cite{ioffe2015batch}, but it can also lead to an overall highest accuracy \cite{ioffe2015batch} and, as stated before, render the network stronger and more robust against overfitting.

\subsection{System Overview}

In this subsection, we analyse the pipeline of our approach. Initially, we operate on clean rest-frame spectral profiles, each consisting of $3750$ bins. These wavelength-related bins correspond to the spectral density flux value of each observation, for that certain wavelength range ($\Delta\lambda$ = $5\AA,\,\lambda$ = $[1252.5, 20002.5]\AA\,$). Our first goal is to create valid redshifted variations using the formula:

\begin{equation}
log(1+z) =\,log(\lambda_{obs})\,-\,log(\lambda_{emit}) \Leftrightarrow 1+z = \frac{\lambda_{obs}}{\lambda_{emit}},
\end{equation}

where $\lambda_{emit}$ is the original, rest-frame wavelength, $z$ is the redshift we want to apply and $\lambda_{obs}$ is the wavelength that will ultimately be observed, for the given redshift value. This formula is linear on logarithmic scale. For the conduction of our experiments, we work on the redshift range of $z = [1,\,1.8)$, which is very similar to what Euclid is expected to detect. Also, to avoid redundant operations and to establish a simpler and a faster network we use a subset of the wavelength range of each redshifted example (instead of the entirety of the available spectrum), based on Euclid's spectroscopic specifications $(1.1 - 2.0 \mu m \Leftrightarrow 11000 - 20000 \AA)$. That means that all the training \& testing observations will be of equal size $\frac{20000 - 11000}{\Delta\lambda} =$ 1800 bins.

For the ``Regression to Classification" transition our working redshift range of $[1,\, 1.8)$ must be split into $800$ non-overlapping, equally-sized slots resulting in a resolution of 0.001, consistent with Euclid expectations. Each slot will correspond to the related ordinal class (from 0 to 799), which in turn must be converted into the 1-Hot Encoding format to match the final predictions procured by the final Softmax Layer of the CNN. A certain real-valued redshift of a given spectral profile will be essentially transformed into the ordinal class that corresponds to the redshift slot it belongs to. Finally, for the predictions, shallower and deeper variations of a Convolutional Neural Network will be trained, with 1,2 $\&$ 3 Convolutional (+ ReLU) Layers, along with a Fully-Connected Layer as the final Classification Layer. 

\section{A Deeper Perspective On The Data}\label{ADeeperPerspective}

The simulated dataset used is modeled after the upcoming Euclid satellite galaxy survey \cite{laureijs2011euclid}. When generating a large, realistic, simulated spectroscopic dataset, we need to ensure that it is representative of the expected quality of the Euclid data. A first requirement is to have a realistic distribution of galaxies in several photometric observational parameters. We want the simulated data to follow representative redshift, color, magnitude and spectral type distributions. These quantities depend on each other in intricate ways, and correctly capturing the correlations is important if we want to have a realistic assessment of the accuracy of our proposed method. To that end, we define a master catalog for the analyses with the COSMOSSNAP simulation pipeline \cite{jouvel2009designing}, which calibrates property distributions with real data from the COSMOS survey~\cite{capak2007first}. The generated COSMOS mock Catalog (CMC) is based on the 30-band COSMOS photometric redshift catalogue with magnitudes, colors, shapes and photometric redshifts for $538.000$ galaxies on an effective area of $1.24 \ deg^ 2$ in the sky, down to an $i$-band magnitude of $\sim 24.5$ \cite{ilbert2008cosmos}. The idea behind the simulation is to convert these real properties into simulated properties. Based on the fluxes of each galaxy, it is possible to select the best-matching SED from a library of predefined spectroscopic templates. With a ``true" redshift and an SED associated to each galaxy, any of their observational properties can then be forward-simulated, ensuring that their properties correspond to what is observed in the real Universe.

For the specific purposes of this analysis, we require realistic SEDs and emission line strengths. Euclid will observe approximately 50 million spectra in the wavelength range $11000 - 20000\,\mathrm{\AA}$ with a mean resolution $R = 250$, where $R =  \frac{\lambda}{\Delta\lambda}$. To obtain realistic spectral templates, we start by selecting a $50\%$ random subset of the galaxies that are below redshift $z=1$ with H$\alpha$ flux above $10^{-16} \,erg\, cm^{-2} \,s^{-1}$, and bring them to rest-frame values ($z=0$). We then resample and integrate the flux of the best-fit SEDs at a resolution of $\Delta\lambda = 5\mathrm{\AA}$. This corresponds to  $R=\frac{\lambda}{\Delta\lambda} = 250$ at an observed wavelength of $11000\,\mathrm{\AA}$, if interpreted in rest-frame wavelength at $z = 2$. For the purpose of our analysis, we will retain this choice, even though it implies higher resolution at larger wavelengths. Lastly, we redshift the SEDs to the expected Euclid range. In the particular case where we wish to vary the number of training samples, we generate more than one copy per rest-frame SED at different random redshifts. We will refer to the resampled, integrated, redshifted SEDs as ``clean spectra" for the rest of the analysis.

\begin{table}[!t]
\caption{Comparison of CPU \& GPU training running time, in 3 different benchmark experiments. In the $1^{st}$ and the $2^{nd}$ experiments, we utilize 40,000 and 400,000 training observations, of the idealistic case, in a CNN with 1 Convolutional Layer. In the $3^{rd}$ case, we deploy 40,000 realistic training examples for the training of a CNN with 3 Convolutional Layers.}\label{table_times}
\begin{center}
	\center
	\begin{tabular}{|c|c|c|c|}
	\hline
	\textbf{Experiment $\#$} & \textbf{CPU Time (per epoch)} & \textbf{GPU Time (per epoch)}\\\hline \hline
	1   & 75 sec.  & 11 sec.  \\ \hline
    2   & 735 sec. & 107 sec. \\ \hline
    3   & 158 sec. & 20 sec.  \\ \hline
 	\end{tabular}
\end{center}
\end{table}

For each clean spectrum above, we generate a matched noisy SED. The required sensitivity of the observations is defined in terms of the significance of the detection of the $H\alpha$ Balmer transition line: an unresolved (i.e. sub-resolution)  $H\alpha$ line of spectral density flux $3 \times 10^{-16} erg\,cm^{-2} s^{-1}$ is to be detected at $3.5\sigma$ above the noise in the measurement. We create the noisy dataset by adding white Gaussian noise such that the significance of the faintest detectable $H\alpha$ line according to the criteria above is $1\sigma$. This does not include all potential source of noise and contamination in Euclid observations, such as dust emission from the galaxy and line confusion from overlapping objects. We do not include these effects as they depend on sky position and galaxy clustering, which are not relevant to the assessment of the efficiency and accuracy of redshift estimation. Our choice of Gaussian noise models other realistic effects of the observations, including noise from sources such as the detector read-out, photon counts and intrinsic galaxy flux variations.

\begin{figure}[!b]
\subfloat{\includegraphics[height=6cm,width=\columnwidth]{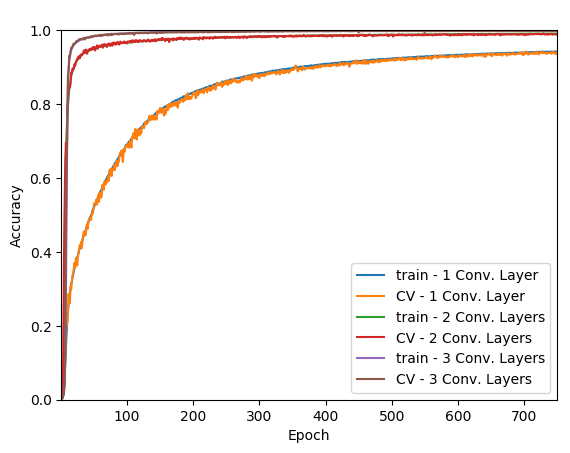}}
\vspace{3mm}
\caption{Accuracy plot, for the Training \& Cross-Validation Sets, for 1,2 \& 3 Convolutional Layers. The x-axis corresponds to the number of executed epochs. In all cases we used the same 400,000 Training Examples.\newline}
\label{layer_depth}
\end{figure}

\section{Experimental Analysis and Discussion}\label{Experimental}

\begin{figure*}[!t]
\centering
\subfloat{\includegraphics[trim=1cm 1cm 1cm 0cm, scale=0.6]{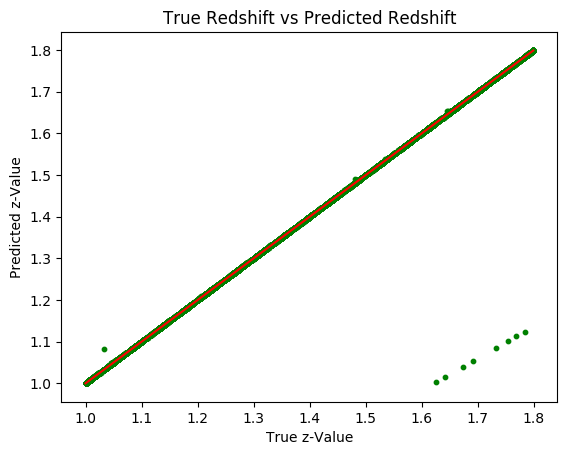}}
\hspace{40pt}
\subfloat{\includegraphics[trim=1cm 1cm 1cm 0cm, scale=0.6]{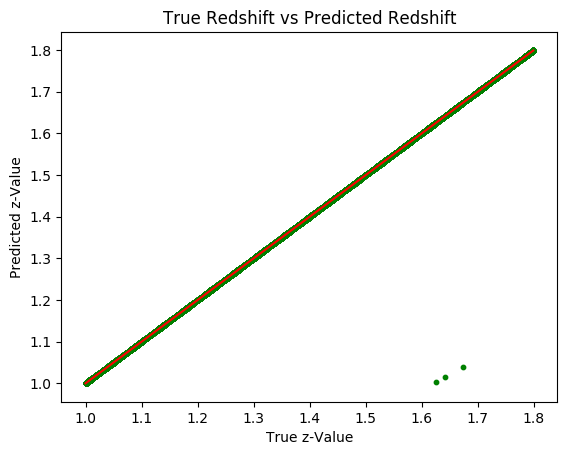}}
\vspace{20pt}
\caption{Classification accuracy achieved by a CNN with one (left) and three (right) Convolutional Layers. The given scatter plots, illustrate points in 2D space that correspond to the true class for each testing observation versus the predicted outcome of the corresponding classifier, for that observation. \newline}
\label{vs_1_3L}
\end{figure*}

We implemented our Deep Learning model with the help of TensorFlow \cite{tensorflow2015-whitepaper} and Keras \cite{chollet2015keras} libraries, in Python code. TensorFlow is an open-source, general-purpose Machine Learning framework for numerical computations, using data flow graphs, developed by Google. Keras is a higher level Deep Learning-specific library, capable of utilizing TensorFlow as a backend engine, with support and frequent updates on most state-of-the-art Deep Learning models and algorithms. Both TensorFlow and Keras have the significant advantage that they can run calculations on GPU, dramatically decreasing the computational time of the network's training, as depicted in Table \ref{table_times}. For the purpose of our experiments we used NVIDIA's GPU model, GeForce GTX 750 Ti. 

As initial pre-experiments have shown, desirable values for the network's different hyperparameters are a kernel size of 8, a number of filters equal to 16 (per convolutional layer) and a stride equal to 1. Additionally, the Adagrad optimizer \cite{duchi2011adaptive} has been used for training, a Gradient Descent-based algorithm with an adaptable learning rate capability, granting the network a bigger flexibility in the learning process and exempting us from the responsibility of tuning an extra hyperparameter.

In both the idealistic and the realistic case, a simple normalization method has been used on all spectral profiles, for compatibility reasons with the CNN, but taking heed, at the same time, that the structure of the data would remain unchanged. The method is depicted in Equation \ref{Eq3}, where $X_{max}$ corresponds to the maximum spectral density flux value encountered in all examples (in absolute terms, given the noisy case) and $X_{original}$ is the initial value for each feature:
\begin{equation}\label{Eq3}
X_{normalized} = \frac{X_{original}}{2 * X_{max}}\, 
\end{equation}

\begin{figure}[!b]
\subfloat{\includegraphics[height=6cm,width=\columnwidth]{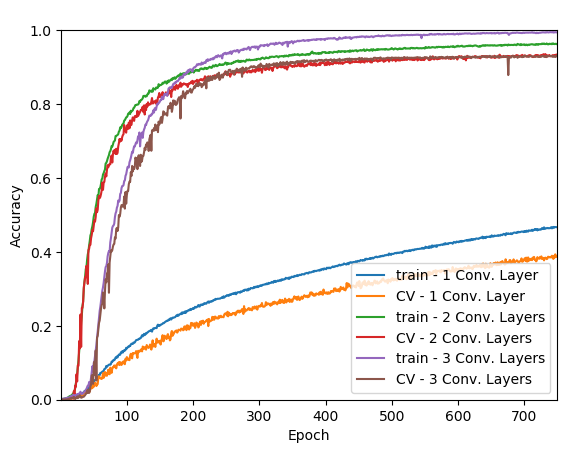}}
\vspace{3mm}
\caption{Training \& Cross-Validation accuracy, for 1,2 \& 3 Convolutional Layers, using a significantly decreased amount of training observations (40,000). Overfitting is introduced, to various extents, based on each case.\newline}
\label{40k_in_train}
\end{figure}

\subsection{Idealistic observations}

\subsubsection{Impact of the Network's Depth}

Our initial experiments revolve around the depth of the Convolutional Neural Network. We have used a fixed number of 400,000 training examples, 10,000 validation and 10,000 testing examples. Our aim is to examine the impact of increasing the depth of the model, on the final outcome. Specifically, we have trained and evaluated CNNs with 1,2 \& 3 Convolutional Layers. In all cases, a final Fully-Connected Layer with $800$ output neurons have been used for classification.

Accuracy is the basic metric that can be used to measure the performance of a trained classifier, during and after the training process. As the training goes by, we expect that the parameters of the network will start to adapt to the problem at hand, thus decreasing the total loss, as defined by the cost function, and, consequently, improving the accuracy percentage. In Figure \ref{layer_depth}, we support this presumption by demonstrating the accuracy's rate of change over the number of training epochs. It can be easily derived that as a CNN becomes deeper, it is clearly more capable to form a reliable solution. Both 2 and 3-layered networks converge very fast and very close to the optimal case, with the latter, narrowly resulting in the best accuracy. On the other hand, the shallowest network is very slow and significantly underperforms compared to the deeper architectures.

More information can be deduced in Figure \ref{vs_1_3L}, where we compare, for the shallowest and for the deepest case, and per testing example, the predicted redshift value outputed by the trained classifier versus the state-of-nature. Ideally, we want all the green dots depicted in each plot to fall upon the diagonal red line that splits the plane in half, meaning that all predicted outcomes coincide with the true values.  As the green dots move farther away from the diagonal, the impact of the faulty predictions become more significant leading to the so called catastrophic outliers. A good estimator is characterized, not only by its ability to procure the best accuracy, but also by its capacity to diminish such irregularities. 

\begin{figure}[!t]
\subfloat{\includegraphics[height=6cm,width=\columnwidth]{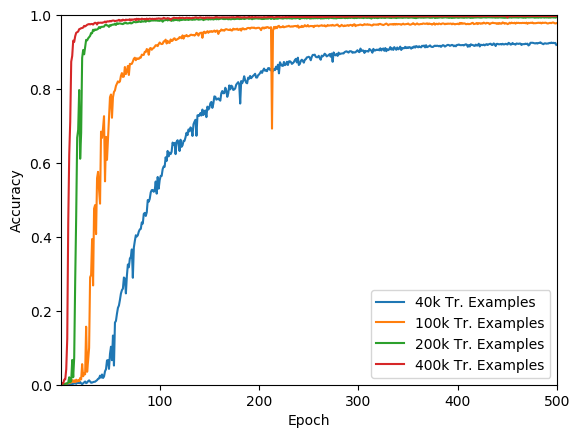}}
\vspace{3mm}
\caption{Validation performance of a 3-layered network, using larger and more limited in size datasets. In all cases the training accuracy (not depicted here) can asymptotically reach 100\% accuracy, after enough epochs.}
\label{varied_data_clean}
\end{figure}

\subsubsection{Data-Driven Analysis}

\begin{figure}[!b]
\subfloat{\includegraphics[height=6cm,width=\columnwidth]{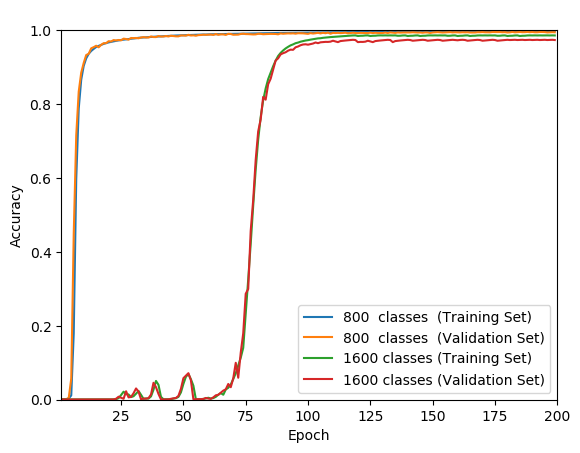}}
\vspace{1mm}
\subfloat{\includegraphics[height=6cm,width=\columnwidth]{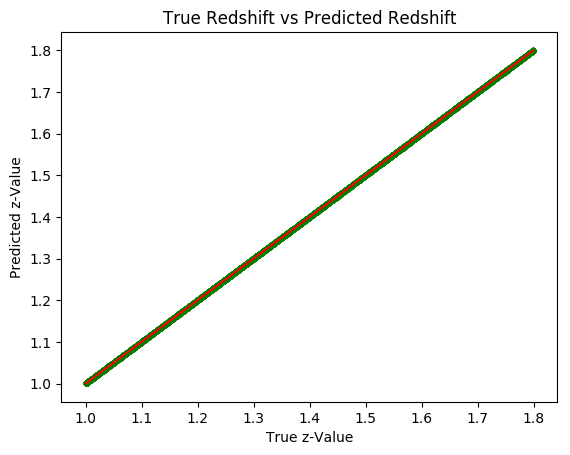}}
\vspace{3mm}
\caption{Performance of a 3-Layered network trained with 400,000 training examples. In the first plot we compare the cases where the redshift estimation problem is transformed into a classification task, with the use of 800 versus 1600 classes. In the second plot, we present the scatter plot of the predicted result versus the state-of-nature of the testing samples, only for the case of 1600 total classes.}
\label{more_classes}
\end{figure}

In this setting, we will explore the significance of broad data availability in the overall performance of the proposed model. As mentioned before, Big Data have revolutionized the way Artificial Neural Networks perform \cite{goodfellow_dlbook}, serving as the main fuel for their conspicuous achievements. Figure \ref{40k_in_train} illustrates the behavior of the same network variations as in previous experiments (1,2 \& 3 Convolutional Layers), using this time a notably more constrained, in size, training set of observations, compared to the previous case. Specifically, we have lowered the number of training examples from 400,000 to 40,000, namely to one-tenth. Compared to the results we have previously examined in Figure \ref{layer_depth}, we can evidently identify a huge gap between the performance of identical models with copious vs more limited amounts of data. It is adequately obvious that in all three cases overfitting is introduced, to various extents, leading to a ``snowball effect", with overoptimistic models that perform well in the training set, but with a decaying performance on the validation and the testing examples.

As a second step, we want to preserve the network's structural and hyperparametric characteristics immutable, whereas altering the amount of training observations utilized in each experimental recurrence. We have deployed a scaling number of training examples beginning from 40,000 observations, then to 100,000 and finally to 200,000 and 400,000 observations and we have used them to train a 3-layered CNN (3 Convolutional + 1 Fully-Connected Layers), in all cases. As shown in Figure \ref{varied_data_clean}, while we increase the exploited amount of data, the curve of the validation accuracy also increases in a smoother and steeper pace, until convergence. On the contrary, when we use less data, the line becomes more unstable, with a delayed convergence and a poorer final performance. It is very important to state, that despite the fact
that the training accuracy can asymptotically reach, in all cases, 100\% accuracy, after enough epochs, the same doesn't apply for the validation accuracy (and respectively for the testing accuracy) with the phenomenon of overfitting taking its toll, mostly in the cases where the volume of the training data is not enough to handle the complexity of the network, failing to generalize in the long term. As we will observe in more detail in the noisy-data case, regularizing techniques, such as Dropout, can actually help battle this phenomenon, but not in a way, that the difference between the training and the validation performance will be completely commensurated.

\subsubsection{Tolerance on Extreme Cases}

\begin{figure}[!t]
\subfloat{\includegraphics[height=6cm,width=\columnwidth]{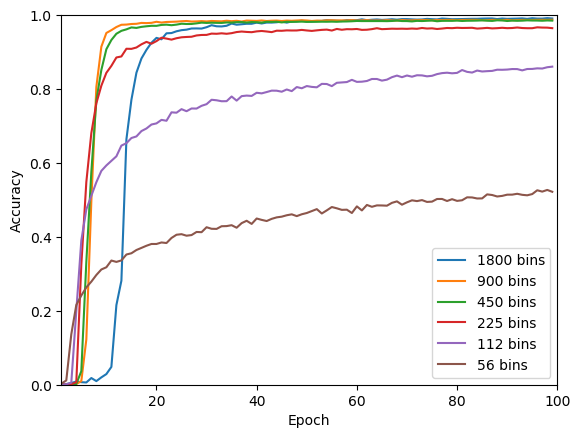}}
\vspace{3mm}
\caption{Validation performance of a 3-Layered network trained with 400,000 training examples. We want to examine the behavior of the model, when trained with data of reduced dimensionality.\newline}
\label{less_bins}
\end{figure}

\begin{figure}[!b]
\subfloat{\includegraphics[height=6cm,width=\columnwidth]{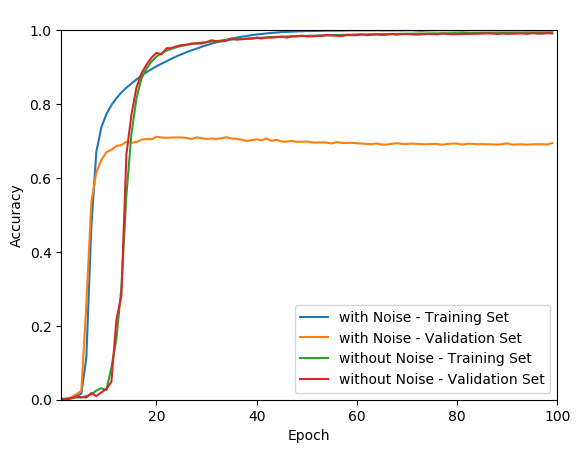}}
\vspace{3mm}
\caption{Comparison of the model's performance, trained with clean and with noisy data (400,000 in both cases). The 3-layered neural network utilizes the same hyperparameters, in both cases, without any form of regularization.\newline}
\label{noisy_vs_clean}
\end{figure}

\begin{figure}[!b]
\subfloat{\includegraphics[height=6cm,width=\columnwidth]{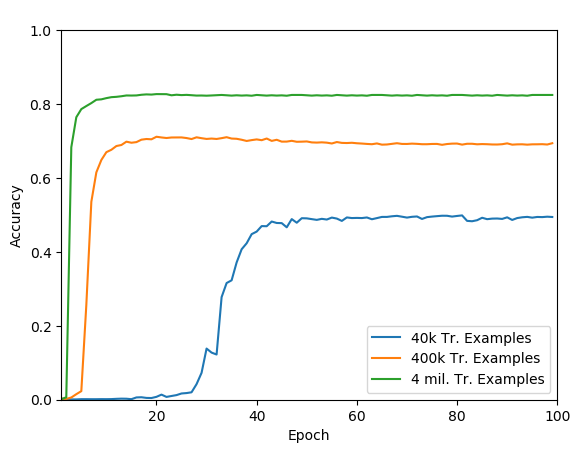}}
\vspace{3mm}
\caption{Accuracy on the Validation Set for different sizes of the Training Set. No regularization has been used.\newline}
\label{big_data_noise}
\end{figure}

\begin{figure*}[!t]
\centering
\subfloat{\includegraphics[trim=1cm 1cm 1cm 0cm, scale=0.6]{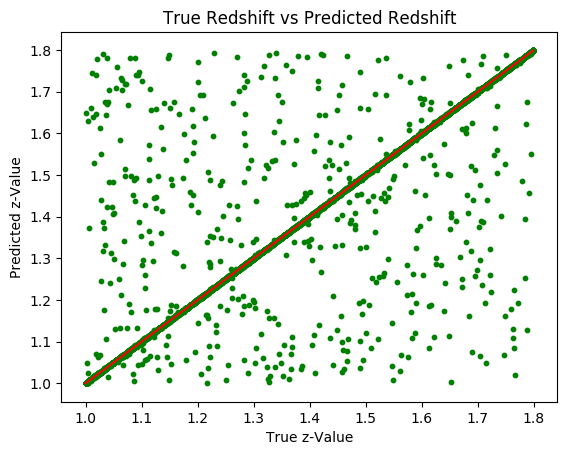}}
\hspace{40pt}
\subfloat{\includegraphics[trim=1cm 1cm 1cm 0cm, scale=0.6]{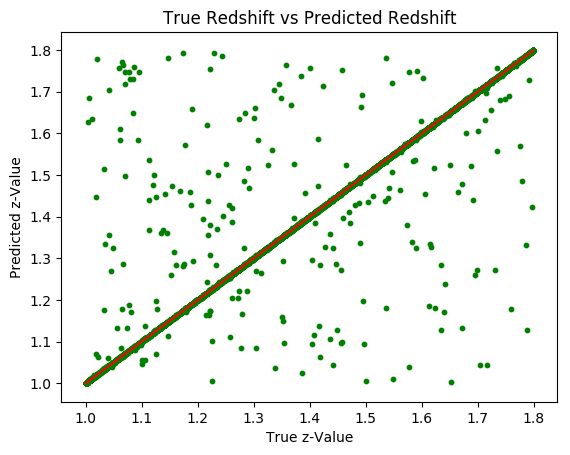}}
\\
\vspace{10pt}
\subfloat{\includegraphics[trim=1cm 1cm 1cm 0cm, scale=0.6]{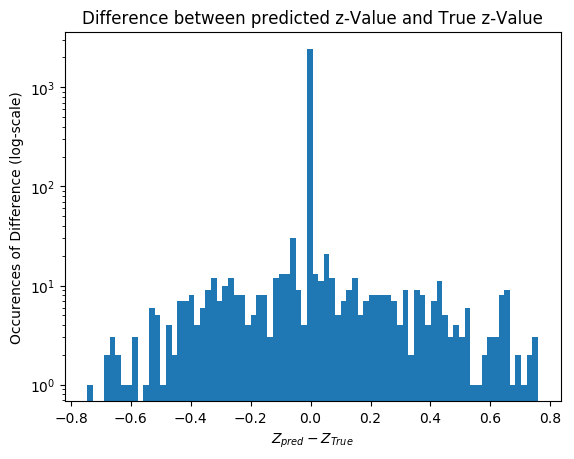}}
\hspace{40pt}
\subfloat{\includegraphics[trim=1cm 1cm 1cm 0cm, scale=0.6]{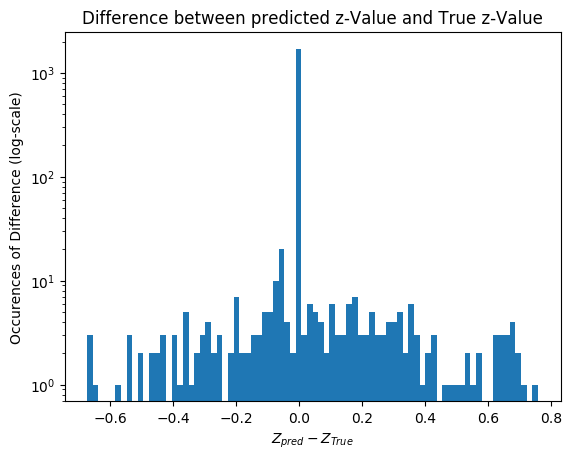}}
\vspace*{7mm}
\caption{Classification scatter plots \& histograms for the realistic case, for 3-layered networks trained with 400,000 Training Examples (column a) \& 4,000,000 Training Examples (column b). The depicted histograms, represent the actual difference in distance (positive or negative) between misclassified estimated values and their corresponding ground truth value versus the frequency of occurrence, in logarithmic scale, for each case.\newline}
\label{extras_noisy}
\end{figure*}

Before advancing to noise-afflicted spectral profiles it is worthsome to investigate some extreme cases, concerning two astrophysical-related aspects of the data. As presented before, one of our main novelties is the realization of the redshift estimation task as a classification task, guided by the specific redshift resolution that Euclid can achieve and leading to the categorization of all possible detectable redshifts into 1 of 800 possible classes. As a first approach, we want to extend our working resolution to a double precision, specifically from 0.001 to 0.0005, meaning that the existing redshift range of [1, 1.8) will be split into 1600 classes instead of 800.

As observed in Figure \ref{more_classes}, doubling the total number of possible classes has a non-critical impact in the predictive capabilities of our approach, given the fact that at convergence, the model produces a similar outcome for the two cases. Despite the fact that doubling the classes leads to a slower convergence, a behavior that can be attributed to the drastical increase of the parameters of the fully-connected layer, the network is still adequate enough to estimate successfully, in the long term, the redshift of new observations. Furthermore, as depicted in the scatter plot of the same figure, we can deduce that increasing the predictive resolution of the CNN, can lead to an increase in the total robustness of the model against catastrophic outliers, given the fact that none of the misclassified observations, in the testing set, exists far from the diagonal red line, namely the optimal error-free case.

In our second approach, we want to challenge the network's predictive capabilities, when presented with lower-dimensional data, and to essentially define which is the turning point, where the abstraction of information becomes more of a strain, rather than a benefit. Having to deal with data that exist in high-dimensional spaces (like in the case of Euclid), can become more of a burden, rather than a blessing, as described by Richard Bellman \cite{bellman2013dynamic}, with the introduction of the very well-known term, of the ``curse of dimensionality". In our case, data dimensionality can be derived by splitting the operating wavelength of the deployed instrument into bins, where each bin corresponds to the spectral density flux value of the wavelength range it describes. Euclid operates in the range of $1.1 - 2.0\,\mu m$ with a bin size of $\Delta\lambda$ = $5\AA$, which implies 1800 different bins, per observation. To reduce that number, we need to increase the wavelength range per bin, by merging it with neighboring cells, namely by adding together their corresponding spectral density flux values. Essentially, we can assert that by lowering the dimensionality of data in this way, we can accomplish to concentrate existing information in cells of compressed knowledge, rather than discarding redundant information.

Figure \ref{less_bins}, actually supports our claim, leading to the conclusion, that, when dealing with clean data, cutting down the number of total wavelength bins into more manageable numbers, can result not only in an congruent performance with the initial model, but also into a faster convergence. On the other hand, oversimplifying the model can be deemed inefficacious, if we take into account the decline of the achieved accuracy in the three low-dimensional cases. A moderate decline in the performance becomes visible in the case of 225 bins, with a more aggressive degeneration of the model in the rest of the cases.

\subsection{Realistic observations}

Having to deal with idealistic data, presumes the ambitious scenario of a reliable denoising technique for the spectra, prior the estimation phase. Although successful methods have been developed in the past \cite{machado2013darth}, \cite{fotiadou2017denoising}, our main aim is to integrate implicitly the denoising operation, in the training of the CNN, meaning that the network should learn to distinguish the noise from the relevant information by itself, without depending on a third party. This way, an autonomous system can be established, with a considerable robustness against noise, a strong feature extractor and essentially a reliable predictive competence. To that end, we have directly used noisy observations, described in Section \ref{ADeeperPerspective}, as the training input of the deployed CNNs.

\begin{figure}[!t]
\subfloat{\includegraphics[height=6cm,width=\columnwidth]{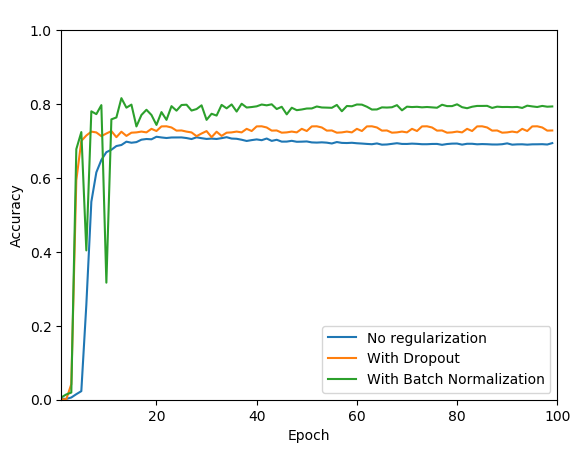}}
\vspace{1mm}
\subfloat{\includegraphics[height=6cm,width=\columnwidth]{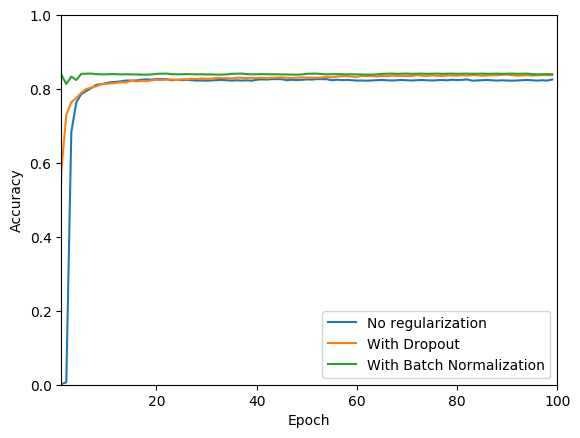}}
\vspace{3mm}
\caption{Impact of regularization, in regard with the size of the training set. In the upper plot, a network trained with 400,000 observations is illustrated, while in the lower plot 4,000,000 training examples have been utilized.}
\label{regularization_in_noise}
\end{figure}

A comparison between the idealistic and the realistic scenarios constitutes the first step, that will lead to an initial realization of the difficulty of our newly set objective. In the illustrated Figure \ref{noisy_vs_clean}, we observe that training a noise-based model with a number of observations that has proven to be sufficient in the clean-based case, leads to an exaggerated performance during the training process, that doesn't apply to newly observed spectra, hence leading to overfitting. Clean data are notably simpler than their noisy counterparts, which in their turn are excessively diverge, meaning that generalization in the latter case is seemingly more difficult. The main intuition to battle this phenomenon lies in drastically increasing the spectral observations used in training. Feeding the network with bigger volumes of data, can mitigate the effects of overfitting, given the fact that despite the network creating a specialized solution fitted for the set of observed spectra, this set tends to become so large that it befits the general case. This intuition is strongly supported by Figure \ref{big_data_noise}, where we compare the performance of similar models, when trained with different-sized sets. Preserving constant hyperparameters and not utilizing any form of regularization, we can derive that, just by increasing in bulk the total amount of data, the network's generalization capabilities also increase in a scalable way. Finally, the new difficulties established by the noisy scenario, also become highly apparent while observing the results of Figure \ref{extras_noisy}. The drastical increase in the number of misclassified samples is more than obvious, subsequently leading to an abrupt rise in the amount and variety of the different catastrophic outliers. Nevertheless, the faulty predictions that lie approximate to the corresponding ground truths, constitute the majority of mispredictions, as verified by the highly populated green mass around the diagonal red line (scatter plots) and the highest bar column bordering the zero value, in the case of the histograms.

\subsubsection{Impact of Regularization}

\begin{figure}[!b]
\subfloat{\includegraphics[height=6cm,width=\columnwidth]{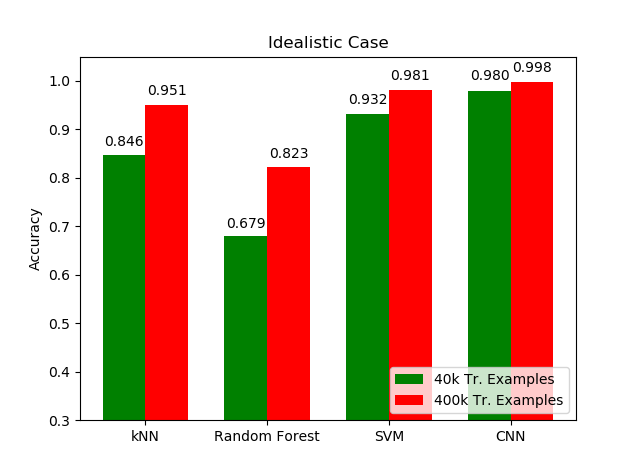}}
\vspace{1mm}
\subfloat{\includegraphics[height=6cm,width=\columnwidth]{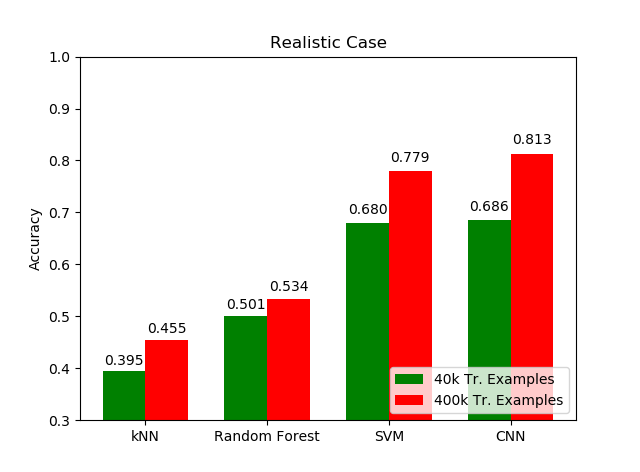}}
\vspace{3mm}
\caption{Comparison bar plots for the k Nearest Neighbours, Random Forest, Support Vector Machines and Convolutional Neural Networks algorithms. We present the best case performance on the test set, for each classifier, in the idealistic and the realistic case, with a limited and an increased amount of training data.\newline}
\label{comparisons}
\end{figure}

\begin{figure*}[!t]
\centering
\subfloat{\includegraphics[trim=1cm 1cm 1cm 0cm, scale=0.4]{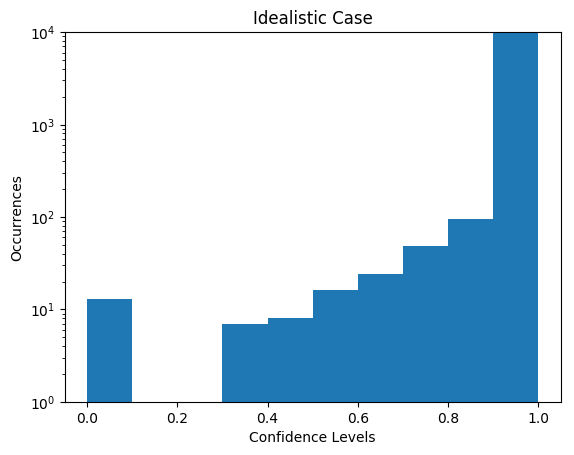}}
\hspace{30pt}
\subfloat{\includegraphics[trim=1cm 1cm 1cm 0cm, scale=0.4]{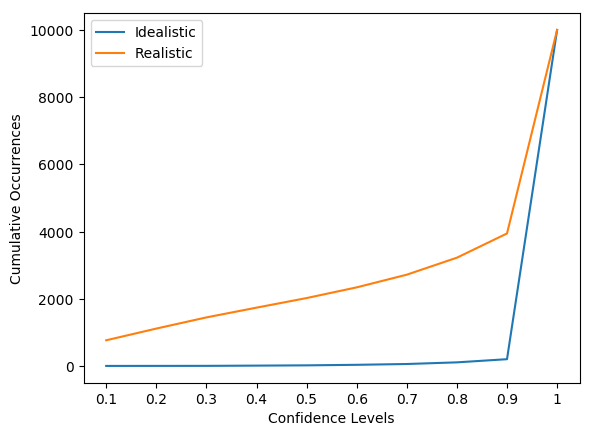}}
\hspace{30pt}
\subfloat{\includegraphics[trim=1cm 1cm 1cm 0cm, scale=0.4]{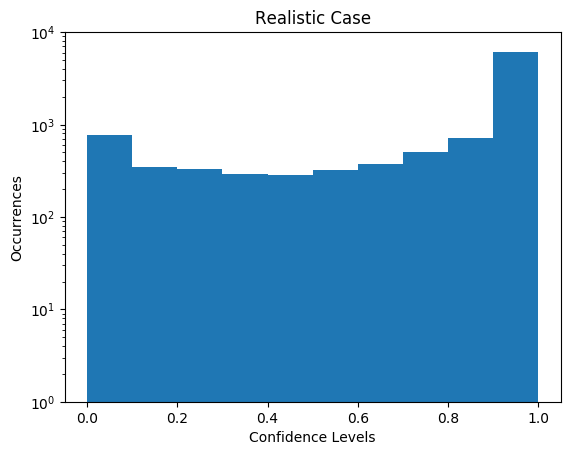}}
\vspace*{7mm}
\caption{Levels of confidence derived by softmax in the testing set. 
The middle plot depicts the cumulative occurrences per level of confidence, for both examined cases. For example, the y-axis value that corresponds to the x-value of 0.4, represents the number of testing observations that obtain a predictive output from the trained model with a confidence that is less than or equal to 0.4. 
The left (idealistic case) and right (realistic case) histograms, exhibit a similar scenario, but not in a cumulative form (and in logarithmic scale).\newline}
\label{probs_conf}
\end{figure*}

The effects of regularization are illustrated in Figure \ref{regularization_in_noise}, in two different settings, one with a Training Set of 400,000 examples and another with a Training Set of 4,000,000 examples. For Batch Normalization, we inserted an extra Batch-Normalization Layer, after each Convolutional Layer (and after ReLU). Although in literature \cite{ioffe2015batch}, the use of Batch Normalization is proposed before the non-linearity, in our case extensive experimental results suggested otherwise. Dropout was introduced only in the Fully-Connected Layer and with a value of $p$ equal to $0.5$, which appeared to yield the best results compared to other cases. It is notable to note that the use of Dropout can be, also, included in the case of the Convolutional Layers, without a mentionable change in the final performance. The number of weights in the Convolutional Layers is dramatically lower compared to the ones in the Fully-Connected Layer, which concentrates the majority of the network's trainable parameters given the large number of output neurons (800 neurons) and the full-connectivity pattern deployed.

As we can see in both examined cases, Dropout can visibly help enhance the network's performance, leading to an increase in the accuracy by \texttildelow{}0.5\% in the worst case and \texttildelow{}1.5\% in the best case. This is not a ground-breaking increase per se, but it is worth mentioning nonetheless. On the other hand, Batch Normalization appears to have a bigger regularizing effect in improving the accuracy of the trained model, yielding a tremendous increase by almost 10\% in the case of 400,000 Training Examples, and a significantly lower gain of \texttildelow{}2\% when trained with 4,000,000 observations. In this final case, even though Batch Normalization still leads to the best performance, its difference compared to Dropout is almost negligible.

\begin{figure*}[!t]
\subfloat[Clean Redshifted Spectral Profile]{\includegraphics[height=4cm,width=5.3cm]{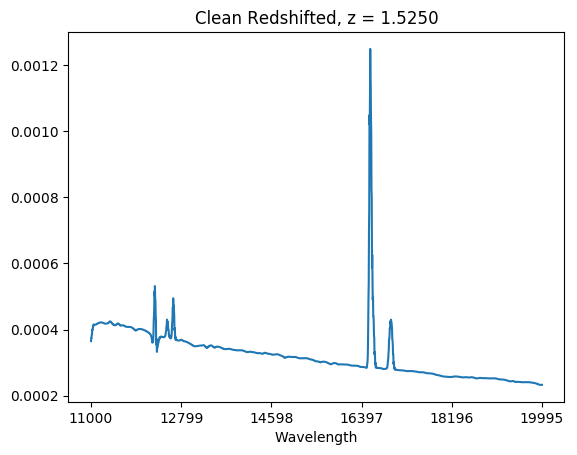}}
\hspace*{1cm}
\subfloat[Activation of $1^{st}$ Conv. Layer]{\includegraphics[height=4cm,width=5.3cm]{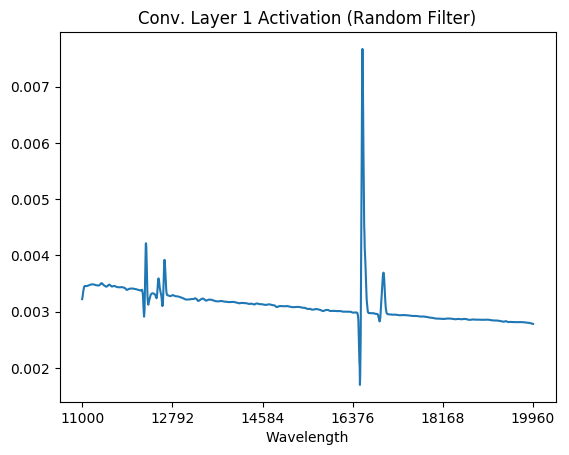}}
\hspace*{1cm}
\subfloat[Activation of $3^{rd}$ Conv. Layer]{\includegraphics[height=4cm,width=5.3cm]{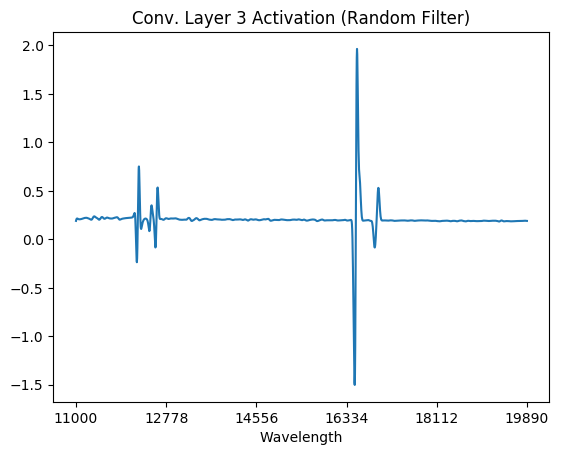}}
\vspace*{7mm}
\caption{A random Testing Example (clean clase) and the corresponding activations of the $1^{st}$ and the $3^{rd}$ Convolutional Layers.\newline}
\label{activations_clean}
\end{figure*}

\begin{figure*}[!t]
\subfloat[Noisy Redshifted Spectral Profile]{\includegraphics[height=4cm,width=5.3cm]{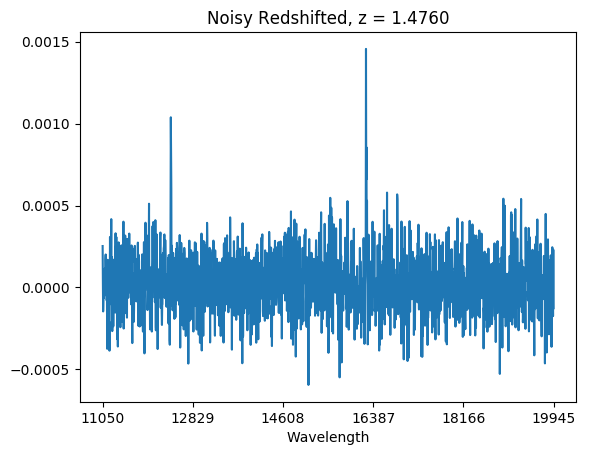}}
\hspace*{1cm}
\subfloat[Activation of $1^{st}$ Conv. Layer]{\includegraphics[height=4cm,width=5.3cm]{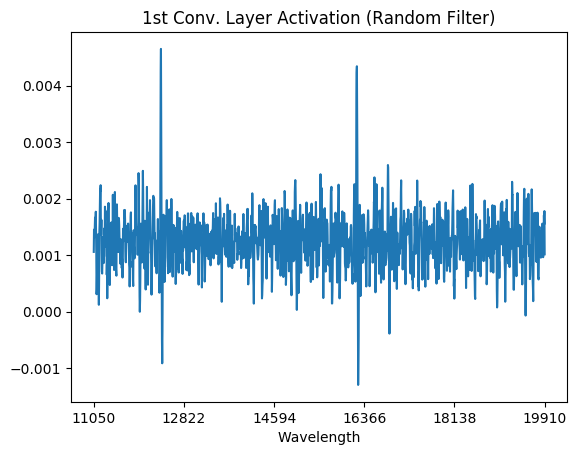}}
\hspace*{1cm}
\subfloat[Activation of $3^{rd}$ Conv. Layer]{\includegraphics[height=4cm,width=5.3cm]{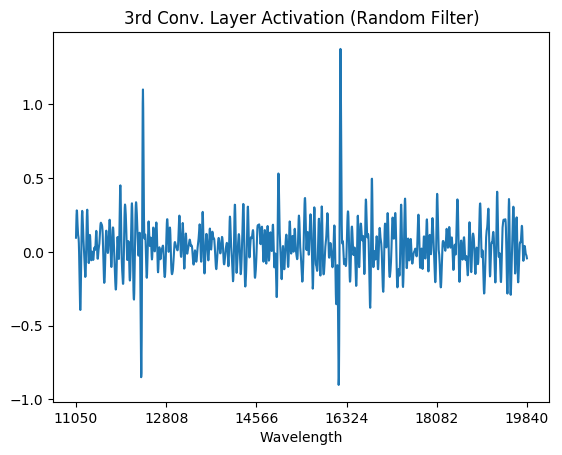}}
\vspace*{7mm}
\caption{A random Testing Example (noisy case) and the corresponding activations of the $1^{st}$ and the $3^{rd}$ Convolutional Layers.\newline}
\label{activations_noisy}
\end{figure*}

\subsection{Comparison With Other Classifiers}

In this subsection, we want to compare the best-case performance of the proposed model, on spectroscopic redshift estimation, against the performance of other popular classifiers, namely k Nearest Neighbours \cite{cover1967nearest}, Random Forests \cite{breiman2001random} and Support Vector Machines \cite{cortes1995support}. The bar plots in Figure \ref{comparisons} corroborate the claim that Convolutional Neural Networks reign supreme as the most effective algorithm for the issue at hand, in all examined cases. The main competitor, in both idealistic and realistic scenarios, stands in the case of the Support Vector Machines (Gaussian kernel), which in our problem is inexpedient to use, given the fact that SVMs are most effective in binary classification scenarios or in cases where the total amount of unique classes is limited. With 800 possible classes to predict, either techniques of ``one-vs-rest" \cite{duda1973pattern} and ``one-vs-one" multiclass classification lead to the need of training 800 and $(800 * 799\, /\, 2) = 319,600$ individual classifiers, accordingly. On the other hand, k Nearest Neighbours and Random Forests  significantly underperform, with a complete failure to cope with the noisy variations of the data, in the realistic case, even with an increased amount of training examples.

\subsection{Levels of Confidence}

As discussed earlier, transforming the redshift estimation problem to a classification procedure provides the benefit of associating each estimation with a level of confidence of the network's certainty that the predicted outcome corresponds to the true redshift value. Using the probabilities produced by the softmax function, we can extract valuable information about the network's robustness, as illustrated in Figure \ref{probs_conf}, where we examine the derived confidence of the best-case trained networks for both idealistic and realistic datasets. In the idealistic scenario, we can observe that the trained model is generally very confident about the validity of its predictions leading to a very steep cumulative curve in the transition from the 90\% to 100\% . As also verified by the corresponding histogram, most of the predictions are associated with a very high probability that lies in the range of (0.9, 1], with a decreased frequency of occurrence as the levels of confidence decrease. This is a very desirable property, given the fact that we want the network to be certain about its designated choice, leading to concrete estimations that are not subject to dispute. In the realistic scenario, although the total confidence of the trained network clearly drops, as expected, still the high confidence choices remain dominant in quantity, compared to the lower cases, which mostly correspond to the misclassified observations.

\subsection{Intermediate Representations}
In this final paragraph, we will briefly examine the undergoing transformation of the input data, as they flow deeper into the network. As previously discussed, Convolutional Neural Networks are excellent feature extractors and can manage to distil important knowledge from raw data, even when suffering from high levels of noise. In Figure \ref{activations_clean}, we can clearly observe that the salient effect of randomly chosen filters from the selected layers, is that as the network deepens, the continuum of the derived intermediate representations is gradually removed, preserving only the characteristic emission and absorption lines of the given spectra (most importantly the $H\alpha$ line). Removing the continuum is one of the key steps that any spectroscopic analysis performs, while on the other hand distinguishing these lines constitutes the key characteristic that will consequently lead to a better discrimination of the different redshift classes. The introduction of mirror amplitudes in the negative half-plane is not of specific importance, given their immediate nullification by the succeeding ReLUs. Furthermore, in the case of the realistic scenario in Figure \ref{activations_noisy}, even though the outright removal of irrelevant information may not be easily achievable, given the low signal-to-noise ratio of the observed spectrum, essentially the network is able to perform a partial denoising of the examined profile, gradually isolating the desired peaks from the faulty discontinuities.

\section{Conclusion}\label{Conclusions}

In this paper, we proposed an alternative solution for the task of spectroscopic redshift estimation, through its transformation from a regression to a classification problem. We deployed a variation of an Artificial Neural Network, commonly known as a Convolutional Neural Network and we thoroughly examined its estimating capabilities for the issue at hand in various settings, using big volumes of training observations that fall into the category of the so called Big Data. Experimental results unveiled the great potential of this radically new approach, in the field of spectroscopic redshift analysis, and triggered the need for a deeper study, concerning Euclid and other spectroscopic surveys. In the case of Euclid, our focus can be concentrated, in the introduction of new noise patterns that will complement the existing noise-scenario to an outright realistic simulation. Using these data, a robust predictive model can be built, capable of pioneering in the area of our study, and a form of transfer learning can be applied \cite{pratt1993discriminability}, exploiting future, real Euclid observations. Another avenue of applications involves other spectroscopic surveys. The Dark Energy Spectroscopic Instrument (DESI) \cite{levi2013desi} is one of the major upcoming cosmological surveys currently under construction and installation in Kitt Peak, Arizona. It will operate in different wavelengths and under different observational and instrumental conditions compared to Euclid, and consequently will be able to detect galaxies with different redshift properties. These two cases will be investigated in our future work.


%

%

\ifCLASSOPTIONcompsoc
  \section*{Acknowledgments}
\else
  \section*{Acknowledgment}
\fi

This work was partially funded by the DEDALE project,
contract no. 665044, within the H2020 Framework Program
of the European Commission.

\ifCLASSOPTIONcaptionsoff
  \newpage
\fi



%

%
%

\bibliographystyle{IEEEtran}
\bibliography{references}{}

%





\end{document}